\documentclass[aps,prb,showpacs,twocolumn,groupedaddress]{revtex4}
\usepackage[dvips]{graphicx}
\def\be{\begin{equation}}
\def\ee{\end{equation}}

\begin{document}
\title{Master equations for pulsed magnetic fields: 
Application to magnetic molecules}
\author{Ioannis Rousochatzakis} 
\email{rusohatz@ameslab.gov}
\author{Marshall Luban}
\email{luban@ameslab.gov}
\affiliation{Department of Physics and Astronomy and Ames Laboratory\\
Iowa State University, Ames, Iowa 50011}

\begin{abstract}
We extend spin-lattice relaxation theory to incorporate the use of pulsed magnetic fields for probing the 
hysteresis effects and magnetization steps and plateaus 
exhibited, at low temperatures, by the dynamical magnetization of magnetic molecules. 
The main assumption made is that the lattice degrees of freedom equilibrate in times much shorter than 
both the experimental time scale (determined by the sweep rate) and the typical spin-lattice relaxation time.
We first consider the isotropic case (a magnetic molecule with a ground state of spin $S$ well separated from 
the excited levels and also the general isotropic Heisenberg Hamiltonian where all energy levels are relevant)
and then we include small off-diagonal terms in the spin Hamiltonian to take into account the
Landau-Zener-St\"{u}ckelberg (LZS) effect. 
In the first case, and for an $S=1/2$ magnetic molecule we arrive at the generalized 
Bloch equation recently used for the magnetic molecule \{V$_6$\} in Phys. Rev. Lett. 94, 147204 (2005). 
An analogous equation is derived
for the magnetization, at low temperatures, of antiferromagnetic ring systems. 
The LZS effect is discussed for magnetic molecules with a low spin ground 
state, for which we arrive at a very convenient set of equations that take into 
account the combined effects of LZS and thermal transitions. 
In particular, these equations explain the deviation from exact magnetization reversal 
at $B\approx 0$ observed in \{V$_6$\}. They also account for the small magnetization plateaus 
(``magnetic Foehn effect''), following the LZS steps, that have been observed in several magnetic molecules. 
Finally, we discuss the role of the Phonon Bottleneck effect at low temperatures and specifically
we indicate how this can give rise to a pronounced Foehn effect.       
\end{abstract}


\maketitle

\section{Introduction}
The subject of magnetic molecules has attracted much attention both for its scientific
importance for studying fundamental issues in nanomagnetism as well as for potential applications. 
Within each molecular unit are embedded a finite number of magnetic ions,  
coupled via Heisenberg super-exchange interactions. Furthermore, the intermolecular magnetic interactions 
are of dipolar origin and can usually be neglected. 
As a result, measurements on crystalline samples reflect the magnetic properties of
isolated individual molecules, with its most prominent feature, arising from the finite number 
of magnetic ions, being the appearance of a discrete magnetic energy level spectrum. 

This feature of the spectrum is reflected in the relaxational behavior 
which mainly arises from the interaction with environmental degrees of freedom, i.e., a ``heat bath'', 
such as phonons: The relaxation times of the dynamical magnetization can become very long 
even at moderately high temperatures. Specifically, in molecules with a high spin 
ground state\cite{Friedman,Sangregorio,Schenker} (single molecule magnets (SMM's)) 
an anisotropy energy barrier is responsible 
for relaxation times as high as $10^{3}-10^{5}$ sec, 
whereas in some molecules with a low spin ground state, they are of the order 
of $10^{-3}-10^0$ sec for $T \lesssim 4$ K.\cite{Rousochatzakis,Nojiri,Chiorescu,Waldmann,Inagaki,Shapira}
The existence of long relaxation times becomes manifest through the appearance of dynamical 
hysteresis effects when using pulsed magnetic fields, 
and this is one of the most exciting phenomena observed in magnetic molecules. 
Clearly, the hysteretic behavior is observable when the experimental time scale $\tau_e$ (determined by the 
field sweep rate) is in the regime of the spin-lattice relaxation times (or shorter). This opportunity is 
available for magnetic 
molecules in conjunction with the current experimental capability 
of using strong magnetic fields with sweep rates as 
high as 1 Tesla/ms.\cite{Rousochatzakis,Nojiri,Shapira,Inagaki}

Another effect manifested in pulsed field measurements, 
is the appearance of abrupt magnetization steps at 
given fields.\cite{Friedman,Sangregorio,Schenker,Rousochatzakis,Nojiri,Shapira,Inagaki} 
These steps are quantum-mechanical in origin and reveal the existence of
small, off-diagonal terms in the spin Hamiltonian which in turn give rise to avoided level crossings in the magnetic 
energy spectrum and to Landau-Zener-St\"{u}ckelberg (LZS)\cite{Landau,Miyashita} transitions. 
The origin of these interactions may be single-ion anisotropy or anisotropic exchange. 
The characteristic energy splitting $\delta$
of the LZS effect varies greatly among magnetic molecules. For istance, in SMM's 
due to the large spin of the ground state, $\delta/k_B$ can be of order $10^{-7}$ K 
(as usual, $k_B$ stands for Boltzmann's constant), whereas in molecules with 
low spin ground state, typically $\delta/k_B \sim 0.1$ K.\cite{Chiorescu,Rousochatzakis} 
This implies that for currently available sweep rates one can probe the non-adiabatic regime in 
SMM's,\cite{Wernsdorfer1} 
whereas in molecules with low spin ground states we are already in 
the extreme adiabatic regime. 
Experimentally, and for the molecules with low spin ground states, there have been observed deviations 
from the pure quantum-mechanical prediction regarding the height of the steps that have been associated 
with the role of dissipation whithin the LZS regime. 

Of particular interest is also another effect induced by dynamic fields, namely 
the appearance of small plateaus\cite{Chiorescu,Shapira,Inagaki} in $M(t)$ following each magnetization 
step, thus giving rise 
to satellite peaks in $dM/dB$. 
It has been first shown\cite{Chiorescu} for the low temperature 
experiments on the magnetic molecule \{ V$_{15}$\} that the origin of these
plateaus is the Phonon Bottleneck (PB) effect.\cite{Chiorescu,Waldmann,Abragam,Stevens}
Numerically solving a quantum master equation that had been previously derived for static 
fields, Saito and Miyashita\cite{Saito} provided an alternative viewpoint of this 
effect, which they termed ``the magnetic Foehn effect'': They suggested that 
this behavior is  widespread whether or not one is in the PB regime and that it is a 
consequence (or an ``after-effect'') of the LZS transitions. 
As it turns out (see Sec. IV below), the PB effect can give rise to an enhancement 
of the Foehn effect. 
This has been recently observed experimentally and will be reported elsewhere.\cite{Wernsdorfer2}  

At present, a first-principles account of such relaxational phenomena in magnetic molecules 
induced by dynamical magnetic fields is lacking.  
Our main goal is to show that one can generalize the conventional 
spin-lattice relaxation theory in the context of pulsed fields studies of magnetic molecules. 
The present work is devoted to (and motivated by) pulsed field studies of molecules 
with a low spin ground state: The simplicity of these systems, apart from providing
a basis for better understanding the main physical ideas, allows one 
to directly compare the predictions of the generalized theory with experimental data. 
Hence, these systems, when subject to pulsed fields, 
provide a convenient means for obtaining information on the various relaxational 
processes and microscopic interactions present in all nanomagnetic systems.  

One such system is the ($S=1/2$) magnetic molecule \{V$_6$\},\cite{Luban,Rousochatzakis} 
which shows both pronounced hysteresis loops as well as 
nearly complete reversals of the magnetization at $B\approx 0$.
The hysteresis loops have been accurately reproduced\cite{Rousochatzakis} using a generalization of the 
standard\cite{Bloch,Slichter,Abragam} Bloch equation which in turn revealed that the one-phonon 
acoustic process is the dominant relaxation mechanism at low temperatures, and in addition
provided an estimate of the spin-phonon coupling energy. 
The first-principles derivation of this equation is provided here within the more general context 
of our analysis. 
On the other hand, the abrupt magnetization reversals at $B\approx 0$
were interpreted as the result of adiabatic LZS transitions 
originating from the existence of a small ($\sim 0.4$ K) intra-molecular, anisotropic exchange. 
The small deviation from the pure quantum-mechanical prediction of complete magnetization reversal
was attributed to dissipation effects inside the LZS regime, but no quantitative account was given
in Ref. \onlinecite{Rousochatzakis}. 
This effect is also analysed in the present work.

Another class of molecules with low spin ground state where the generalized theory can be 
easily applied is that of the antiferromagnetic (AFM) ring systems at low $T$. 
These are magnetic molecules 
comprising an even number of uniformly spaced metal ions arranged as a planar ring (see for example 
Refs. \onlinecite{Fe2,Fe6,Fe10,Fe12,Cr8}). The AFM exchange interactions give rise, 
to a non-magnetic $S=0$ ground state, a first excited $S=1$ triplet state, etc. 
In addition to their hysteretic behavior, these systems can show 
several magnetization steps and sometimes the small magnetization plateaus mentioned above.\cite{Shapira,Inagaki} 
As we show below, the present work accounts for these dynamical effects in a general way.

The organization of this paper is the following.
In Sec. II we develop the spin-lattice relaxation theory in pulsed fields 
for magnetic molecules with a spin $S$ ground state that is well separated
from the excited levels. We arrive at a generalization of the standard master equations 
and show how these lead to the generalized Bloch equation for the case of $S=1/2$ mentioned above.
In Sec. III we extend this theory to include the general isotropic Heisenberg Hamiltonian where all energy 
levels are relevant. We apply this to AFM rings at low $T$
and for fields around the first level-crossing field value, where they behave as two-level systems.
We provide the treatment of dissipative LZS transitions in Sec. IV.  
This is done for the case of a level anti-crossing between two levels with 
different magnetic quantum numbers. We apply the resulting theory 
to the spin 1/2 case and that of AFM rings at low $T$, and 
demonstrate how this theory accounts for 
the deviation from the quantum-mechanical prediction for the magnetization steps as well as the 
formation of the plateaus mentioned above (Foehn effect). 
In this way, in particular, we provide an explanation for the deviation from 
exact magnetization reversal observed in \{V$_6$\}.\cite{Rousochatzakis} 
We also indicate the role of the PB effect at low $T$, and specifically 
how it can give rise to a pronounced Foehn effect. 
Finally, in Sec. V, we provide a brief discussion and summary of the present work.   

\section{master equations for spin $S$}
We first consider a magnetic molecule with a ground state 
of definite total spin $S$, well separated from the excited levels.\footnote{The same theory can be applied
to independent paramagnetic ions with total spin $S$ (see Ref. \onlinecite{PhysicaB}).} 
We assume that the temperature is low enough and 
the field regime covered is well below the lowest level-crossing field, 
so that we only need to consider the ground state spin $S$ level.
We focus on the isotropic case i.e., we do not consider non-diagonal terms in the spin
Hamiltonian. 

We employ the standard\cite{Blum,Breuer,Carmichael} method of treating both the spin and the bath 
degrees of freedom quantum-mechanically.
The Hamiltonian of the combined system (spin $S$ + heat bath) is written as 
\be
H(t)=H_{s}(t) + H_{B} + V~,
\label{10}
\ee
where $H_{s}(t)=\hbar \gamma B(t)S_{z}\equiv \hbar f(t)S_{z}$ corresponds to the Zeeman energy, 
$H_{B}$ is the bath Hamiltonian, and $V$ is the spin-bath coupling. 
As usual, $\gamma=g\mu_B/\hbar$ denotes the electronic gyromagnetic ratio. 
In our notation, $f(t)=\gamma B(t)$ has units of frequency and 
the spin operators are taken to be dimensionless.
Typical sweep forms are shown in figures below; the experimental 
sweep time $\tau_e$ can be as short as 1 ms.

For our purposes it is unnecessary to specify the detailed form of the interaction $V$, however, 
it can always be written in the general form
\be
V=\hbar\sum_{q}A_{q} \otimes R_{q}~,
\label{20}
\ee
where $A_{q}~(R_{q})$ are hermitian operators of the spin (bath) system. This coupling may, for example, 
originate from the modulation of the exchange
coupling terms (between the individual magnetic ions of the molecule) or the modulation of 
the interaction of each individual ion with its local environment (e.g. the crystal field)
which dynamically affects the spin degrees of freedom through the spin-orbit coupling.\cite{Stevens,Mattuck} 
In both cases, the modulation originates from lattice deformations, i.e., phonons. The terms bath and 
environmental degrees of freedom will be understood to mean phonons. 

The equation of motion for the density matrix $\rho_{tot}(t)$ of the combined system follows the von Neumann 
equation\cite{Blum,Breuer,Carmichael}  
\be
\dot{\rho}_{tot}(t)=-i[H(t)/\hbar,\rho_{tot}(t)]~.
\label{30}
\ee 
We switch to the interaction picture 
\be
\widetilde{\rho}_{tot}=e^{i(F(t)S_{z}+H_{B}t/\hbar)}~\rho_{tot}~e ^{-i(F(t)S_{z}+H_{B}t/\hbar)}~,
\label{40}
\ee
where $F(t)\equiv \int_{0}^{t}~f(t')~dt'$. The equation satisfied by $\widetilde{\rho}_{tot}$ is
\be
\dot{\widetilde{\rho}}_{tot}(t)=-i[\widetilde{V}(t)/\hbar,\widetilde{\rho}_{tot}(t)]~,
\label{50}
\ee
where we have defined 
\be
\widetilde{V}(t)=\hbar \sum_{q} \widetilde{A}_{q}(t) \otimes \widetilde{R}_{q}(t),
\label{60}
\ee
with $\widetilde{A}_{q}(t)=e^{iF(t)S_{z}}~A_{q}~e^{-iF(t)S_{z}}$, and similarly 
$\widetilde{R}_{q}(t)=e^{iH_{B}t/\hbar}~R_{q}~e^{-iH_{B}t/\hbar}$.
Now, all the relevant information about the spin system is contained in the so-called reduced 
density matrix $\rho(t) \equiv \textrm{Tr}_{b}~(\rho_{tot}(t))$, since for example 
\be
<S_z>=\textrm{Tr}_s\textrm{Tr}_{b}\{\rho_{tot}(t) S_z\}=\textrm{Tr}_s\{\rho(t) S_{z}\}~,
\label{70}
\ee
where Tr$_{s}$ (Tr$_{b}$), denotes the partial trace over the spin (bath) degrees of freedom.
Thus we are mainly interested in finding an equation of motion for $\rho(t)$.
We first make the assumption that the spin-bath coupling is sufficiently weak 
so that the Born approximation can be used. 
Furthermore, we consider temperatures high enough ($T > 1$ K) so that 
the number of available phonon modes per spin is very large and as a result 
the phonons equilibrate \emph{independently} from the spins. 
Given this consideration and the expectation that the phonon relaxation times $\tau_b$ 
(typically $\tau_b < 10^{-6}$ sec) are much shorter than both the experimental time scale $\tau_e$ and 
the spin-lattice relaxation time $\tau_s$,
the density matrix of the compined system (spin system + bath) 
can be factored as $\widetilde{\rho}_{tot}(t)\approx\widetilde{\rho}(t) \otimes \rho_{B}$, 
where $\rho_{B}=e^{-\beta H_{B}}/Z_{B}$ describes the stationary state of the heat bath at temperature $T$.
Here $Z_{B}$ denotes the bath partition function, and $\beta\equiv 1/(k_{B}T)$.
 The above factorization of the total density matrix  
is expected to break down at sufficiently low temperatures (typically $T<1$ K) where one 
expects the phonon bottleneck (PB) effect\cite{Chiorescu,Abragam,Stevens} to take place. 
Employing the above approximations, one arrives at the following integro-differential equation of motion 
for $\widetilde{\rho}$,\cite{Blum,Breuer,Carmichael}
\be
\dot{\widetilde{\rho}}(t)=\frac{1}{\hbar^{2}} \int_{0}^{t} du~\textrm{Tr}_{b} 
[[\widetilde{V}(u),\widetilde{\rho}(u) \otimes \rho_{B}],\widetilde{V}(t)]~.
\label{80}
\ee  
We denote the adiabatic eigenvalues of $H_{s}(t)$ by $\epsilon_{M}(t)=\hbar f(t) M$, where $M=-S,\ldots,S$. 
The adiabatic excitation frequencies of $H_{s}(t)$ are of the form $\omega_{\mu}(t)=f(t)\mu$, where
$\mu=-2S,\ldots,2S$. 
Now, for a given operator $A_{a}$ of the spin system it is very convenient to construct the so-called 
eigenoperator $A_{a,\mu}$ corresponding to a given excitation frequency $\omega_{\mu}(t)$ as
\begin{eqnarray}
A_{q,\mu}\equiv \sum_{M,M'} (A_q)_{MM'}|M><M'|~\delta_{M'-M, \mu}~,
\label{90}
\end{eqnarray}
where $\delta_{i j}$ denotes the Kronecker delta symbol. 
These operators obey the equation 
\be
A_{q}=\sum_{\mu} A_{q,\mu}~,
\label{100}
\ee
where the sum extends over all possible integers $\mu$. The reason for introducing the 
eigenoperators $A_{q,\mu}$ is that they take a very simple form in our interaction picture, namely
\be
\widetilde{A}_{q,\mu}(t)=e^{-iF(t)\mu}~A_{q,\mu}~.
\label{110}
\ee
We note in particular that for a static external field $B_{0}$, we have $F(t)=\omega_{0} t$, 
where $\omega_{0}\equiv \gamma B_{0}$ and the phase factor in Eq. (\ref{110}) becomes the 
familiar form $\textrm{exp}(-i \omega_{0} t)$.
In the present case, Eq. (\ref{80}) along with Eqs. (\ref{60}), (\ref{100}) and (\ref{110}) 
and the variable change $u\rightarrow t-u$, give
\begin{eqnarray}
\dot{\widetilde{\rho}}(t)=\sum_{qq',\mu\mu'}\int_{0}^{t}du 
e^{-iF(t)\mu'}e^{-iF(t-u)\mu} (A_{q',\mu}\widetilde{\rho}(t-u)A_{q,\mu'}\nonumber\\
-A_{q,\mu'}A_{q',\mu}\widetilde{\rho}(t-u)) 
<\widetilde{R}_{q}(u) \widetilde{R}_{q'}(0)>_{B} + \emph{h.c.},
\label{120}
\end{eqnarray}
where the quantities in angular brackets 
\be
<\widetilde{R}_{q}(u) \widetilde{R}_{q'}(0)>_{B} \equiv 
\textrm{Tr}_{b} (\rho_{B} \widetilde{R}_{q}(u) \widetilde{R}_{q'}(0))~, 
\label{130}
\ee
are equilibrium time correlation functions of the bath, and the symbol \emph{h.c.} denotes hermitian conjugate.
As mentioned already, $\tau_b<<\tau_s,\tau_e$. This allows one to perform the following simplifications.
First, we may extend the upper limit of integration in Eq. (\ref{120}) to infinity. Second, the variation
of $\widetilde{\rho}(t-u)$ in the time scale of $\tau_b$ is extremely small (since $\tau_s >> \tau_b$), 
so it is justified to perform the usual Markov approximation,\cite{Blum,Breuer,Carmichael} namely we replace 
$\widetilde{\rho}(t-u)$ by $\widetilde{\rho}(t)$. 
Similarly, the variation of $F(t-u)$ in the time scale of $\tau_b$ is also small. 
Therefore we may approximate $F(t-u)$ by 
\be
F(t-u)\approx F(t)-u\cdot \dot{F}(t)=F(t)-uf(t)~,
\label{140}
\ee
i.e., by a Taylor expansion through first order in $u$. The neglect of higher order terms 
is valid since $u^{2}\dot{f}(t)<<uf(t)$, is equivalent to 
$\tau_b << \tau_e$ (with $\tau_b$ taken 
as the maximum value of $u$, and $\tau_e$ as a typical value of $f/\dot{f}$). 
Substituting Eq. (\ref{140}) into Eq. (\ref{120}), we obtain
\begin{eqnarray}
\dot{\widetilde{\rho}}(t)=\sum_{qq',\mu\mu'} e^{-iF(t)(\mu'+\mu)} 
\Gamma_{qq'}(\omega_{\mu}(t))\times\nonumber\\
(A_{q',\mu}\widetilde{\rho}(t)A_{q,\mu'}-A_{q,\mu'}A_{q',\mu}\widetilde{\rho}(t)) + \emph{h.c.}~,
\label{150}
\end{eqnarray}
where we define the bath correlation functions, at the time-dependent frequency $\omega_{\mu}(t)$, 
as
\be
\Gamma_{qq'}(\omega_{\mu}(t))\equiv\int_{0}^{\infty}du~e^{i\omega_{\mu}(t)~u} 
<\widetilde{R}_{q}(u)\widetilde{R}_{q'}(0)>_{B}.
\label{160}
\ee
We emphasize that the `adiabatic' $\omega_{\mu}(t)$ factor in Eq. (\ref{160}) 
originates from the second term in the 
Taylor expansion of $F(t-u)$ and it has important implications in what follows.
In particular, since $\omega_{\mu}(t)$ is proportional to the instantaneous field $B(t)$, it will lead to 
an equation of motion with relaxation rates that depend explicitly on $B(t)$.

We further adopt the so-called rotating wave approximation (RWA):\cite{Blum,Breuer,Carmichael}
The relative rate of change of a typical phase factor of Eq. (\ref{150}), with $\mu + \mu'\ne 0$ 
is proportional to $\gamma B(t)$ which is of order $10^{11}$ s$^{-1}$ (for $B\sim1$ Tesla). This implies that 
such non-secular terms ``oscillate'' very rapidly during the experimental 
time scale, and thus do not appreciably contribute to the dynamics 
in Eq. (\ref{150}).\footnote{This holds true as long as $B(t)$ remains nonzero; 
in the level-crossing regime, $B(t)\approx 0$, the RWA is not strictly valid. However, 
the time interval $\delta t$ over which the non-secular terms cannot be neglected
is extremely small compared to the experimental time scale: For $\tau_e \sim 1$ ms one gets
$\delta t\sim 10^{-8}$ ms and therefore, given the much slower relaxation rates $\tau_s$, the inclusion
of these terms does not change anything in such a small time interval.         
Hence, the RWA can be safely used for all fields, even the level-crossing regime.} 
Thus, by retaining only the terms with $\mu'=-\mu$, we may replace Eq. (\ref{150}) by 
\begin{eqnarray}
\dot{\widetilde{\rho}}(t)=\sum_{qq',\mu}\Gamma_{qq'}(\omega_{\mu}(t))
(A_{q',\mu} \widetilde{\rho}(t)A_{q,-\mu}\nonumber\\ 
-A_{q,-\mu}A_{q',\mu}\widetilde{\rho}(t)) + \emph{h.c.}~.
\label{180}
\end{eqnarray}
This is our generalized master equation in the weak coupling and RWA limit and for slowly changing 
(compared to $\tau_{b}$) external fields.

Before we discuss the major consequences of this generalized master equation, we go one step further and 
derive the equations of motion for the populations $\rho_{MM}$ of the various states $|M>$.  
Using the matrix elements
\be
(A_{q,\mu})_{MM'}=(A_{q})_{MM'} \delta_{M'-M,\mu},
\label{190}
\ee 
and the relation $\widetilde{\rho}_{MM}=\rho_{MM}$, one finds that the populations 
decouple from the non-diagonal terms  
and evolve according to the following generalization of the standard Pauli master equation
\be
\dot{\rho}_{MM} = \sum_{M'} W_{M'M}(t) \rho_{M'M'} 
- \sum_{M'} W_{MM'}(t) \rho_{MM}~.
\label{200} 
\ee
The transition rates $W_{M\rightarrow M'}\equiv W_{MM'}$ are defined as
\be
W_{MM'}=\sum_{q,q'}\gamma_{qq'}(\omega_{MM'}(t))(A_{q})_{MM'}(A_{q'})_{M'M},
\label{210}
\ee
where $\hbar \omega_{MM'}\equiv \epsilon_{M}-\epsilon_{M'}$, and
\begin{eqnarray}
\gamma_{qq'}(\omega)\equiv \Gamma_{qq'}(\omega) + \Gamma_{q'q}^{*}(\omega)=\nonumber\\
=\int_{-\infty}^{+\infty} du e^{i\omega u} <\widetilde{R}_{q}(u) \widetilde{R}_{q'}(0)>_{B}~.
\label{220}
\end{eqnarray}
An important property of the transition rates is the detailed balance condition that 
arises from the following quantum property\cite{Blum,Breuer,Carmichael} of the (stationary) bath correlation 
functions appearing in Eq. (\ref{220})
\be
\gamma_{qq'}(-\omega)= \textrm{exp}(-\beta \hbar \omega ) \gamma_{q'q}(\omega)~.
\label{230}
\ee
The detailed balance condition follows straightforwardly from Eq. (\ref{210}) along with Eq. (\ref{230})
\be
W_{M'M}=\textrm{exp}(-\beta \hbar \omega_{MM'})W_{MM'}~.
\label{240}
\ee
We have now arrived at the main results of this Section. 
A comparison with the standard relaxation theory (for static fields) 
shows clearly the physics underlying the generalization made here: 
Equation (\ref{210}) involves transition rates that depend on the adiabatic 
energy excitations of the spin system, which in turn 
are proportional to $B(t)$. All information about specific details of relaxation mechanisms is contained in 
the bath correlation functions as well as the matrix elements $(A_q)_{MM'}$ (see Eq. (\ref{210})). 

Reviewing our derivation, the key assumption made is that the bath degrees of freedom equilibrate \emph{independently}
from the spins in times $\tau_b << \tau_e, \tau_{s}$. The basis of this assumption is that the 
number of available environmental modes per spin is so large that the bath can be considered to be a 
large reservoir, hence allowing the neglect of any feedback from the spin dynamics. This allowed the decomposition 
$\rho_{tot}(t)\approx \rho(t)\otimes \rho_B$ and also the Taylor expansion (particularly the second term) 
in Eq. (\ref{140}). Despite the intuitive appeal of this assumption, it must be checked by comparing
the predictions of the resulting theory with experimental data. 

A simple, realistic system where the present theory is easily applicable and is 
in fact in excellent agreement with experimental data
(for $T > 1.5$ K) is the magnetic molecule \{V$_6$\},\cite{Rousochatzakis} mentioned in the Introduction. 
It is straightforward to show that the generalized Bloch equation used in Ref. \onlinecite{Rousochatzakis}, 
to reproduce the experimental data for this $S=1/2$ system follows immediately 
from the present theoretical framework: 
For $S=1/2$, the equation of motion (Eq. (\ref{200})) for the populations of the spin-up ($|+>$) and 
spin-down ($|->$) states reads
\begin{eqnarray}
\dot{\rho}_{++}=W_{-+}(t) \rho_{--} - W_{+-}(t) \rho_{++}~,\nonumber\\
\dot{\rho}_{--}=W_{+-}(t) \rho_{++} - W_{-+}(t) \rho_{--}~.
\label{250}
\end{eqnarray}
On using the normalization condition $\rho_{++}+\rho_{--}=1$,
and the detailed balance condition $W_{+-}=\textrm{exp}(\beta \hbar \Omega) W_{-+}$ 
(here $\Omega\equiv 2\gamma B$) one finds 
that the magnetic moment per spin, $M\equiv -\hbar\gamma <S_z>$, follows the 
equation
\be
\dot{M}(t)=\frac{1}{\tau_{s}(T,B(t))} \left(M_{eq}(T,B(t))-M(t)\right)~,
\label{270}
\ee
where $M_{eq}=(\hbar\gamma/2)\tanh\left(\beta \hbar \Omega/2\right)$ and
and $1/\tau_{s}=W_{-+} \left(1+\textrm{exp}(\beta\hbar\Omega)\right)$.
This is the generalized Bloch equation that
was used in Ref. \onlinecite{Rousochatzakis}. 
Its physical interpretation is that $M(t)$
relaxes towards the instantaneous equilibrium value $M_{eq}(T,B(t))$ with a relaxation rate that depends 
explicitly on $B(t)$.  

In principle one can invert Eq. (\ref{270}) and extract $1/\tau_{s}$ in terms of $M(t)$ and $\dot{M}(t)$ 
obtained by experiment and the adiabatic equilibrium magnetization $M_{eq}(T,B(t))$.
Alternatively, one can directly compare the experimental data 
with a numerical solution of Eq. (\ref{270}) by choosing a physically appropriate functional form 
of $1/\tau_{s}(T,B(t))$ and adjusting the free parameters. Due to the explicit dependence 
of $1/\tau_{s}$ and $M_{eq}$ on $B(t)$ one can obtain information on the underlying specific relaxation mechanism(s) 
by using different sweep forms $B(t)$. 
Along these lines, it was confirmed in Ref. \onlinecite{Rousochatzakis} that for the magnetic molecule \{V$_6$\} 
and for $1<T<5$ K, the dominant contribution to $1/\tau_s$ is the one-phonon processes term $1/\tau_s^*$ given by
\be
1/\tau_s^*=A \Omega(t)^{3} \coth\left(\beta\hbar\Omega(t)/2\right)~,
\label{300}
\ee
with $A=3\textrm{V}_{sl}^2/(2\pi\hbar\rho v^5)$, where $v$ denotes the sound velocity, 
$\rho$ the mass density, and V$_{sl}$ the characteristic energy modulation of the given spin-phonon 
coupling mechanism.\cite{Abragam,Stevens} Apart from the establishment of 
the dominant relaxation mechanism at $1<T<5$ K, 
a first estimate of V$_{sl}$ ($\sim 0.35$ K) was obtained.
More generally, the excellent agreement of this theory with experimental data signifies that our 
starting assumptions are valid for \{V$_6$\} for $T > 1$ K. 
However, our main assumption, namely that the phonons remain in equilibrium at all experimental times, 
can be expected to break down at lower $T$ since the number of 
available resonant phonons per molecule rapidly decreases on cooling, and the phonon bottleneck (PB) 
effect\cite{Chiorescu,Abragam,Stevens} takes place. In fact, preliminary data\cite{Ajiro} 
for \{V$_6$\} at $T=0.6$ K shows a significant deviation from the theory suggesting the onset of the PB effect. 

We remark that the theory of this Section, which is based on the Hamiltonian 
of Eq. (\ref{10}), cannot account for the magnetization steps observed 
in \{V$_6$\} at the level-crossing regime $B\approx 0$. 
These, however, can in fact be explained in terms of adiabatic 
LZS transitions. The necessary extension of our theory is given in Sec. IV. 
    
\section{Master equations for the general isotropic Heisenberg model}
Here we extend the previous analysis and discuss the general isotropic Heisenberg model
where all the energy levels are relevant. 
The analysis is parallel to the above and straightforward, 
and thus only the main new points are emphasized.  

The Hamiltonian of the combined system (magnetic molecule + heat bath) is again given by Eq. (\ref{10})
where now $H_s(t)$ explicitly includes the Heisenberg exchange Hamiltonian
\be
H_{0}=\sum_{i<j}J_{ij}\vec{S}_{i}\cdot\vec{S}_{j}~,
\label{320}
\ee   
and where $J_{ij}$ denotes the exchange constants between the spins at sites $i$ and $j$.  
The eigenstates $|n>$ of $H_s$ are of the form $|n>=|\nu~S~M>$, where $\nu$ corresponds to 
additional quantum numbers. 
Since $H_{0}$ and $S_{z}$ commute one can define the interaction picture by
\be
\widetilde{\rho}=e^{i(H_0t/\hbar+F(t)S_z)}~\rho~e^{-i(H_0t/\hbar+F(t)S_z)}.
\label{330}
\ee  
In addition, the adiabatic energy eigenvalues have the form $\epsilon_n(t)=\epsilon_{0n}+\hbar f(t) M$, 
where the first term corresponds to the zero-field spectrum of the exchange Hamiltonian (Eq. (\ref{320})) 
and the second term to the Zeeman splitting energy. Hence,
the adiabatic excitation frequencies are of the form $\omega(t)=\omega_{0}+f(t)\mu$, where
$\hbar\omega_{0}\equiv \epsilon_{0n'}-\epsilon_{0n}$ and $\mu=M'-M$. 
Thus, any given excitation frequency $\omega(t)$ can be characterized completely by $\omega_{0}$ 
and $\mu$. As before, we introduce a set of eigenoperator $A_{q}(\omega_{0},\mu)$ given by
\begin{eqnarray}
A_{q}(\omega_{0},\mu)\equiv \sum_{n,n'} (A_q)_{nn'}|n> <n'|\times\nonumber\\
\delta_{M'-M,\mu}~\delta(\epsilon_{0n'}-\epsilon_{0n},\hbar \omega_{0})~,
\label{340}
\end{eqnarray}
which take the following form in the interaction picture,
\be
\widetilde{A}_{q}(\omega_{0},\mu)=e^{-i\omega_{0}t} e^{-iF(t)\mu} A_{q}(\omega_{0},\mu).
\label{350}
\ee
Employing the same steps as in Sec. II, one arrives at the master equation
\begin{eqnarray}
\dot{\widetilde{\rho}}(t)=\sum_{qq'}\sum_{\omega(t)} \Gamma_{qq'}(\omega(t))
(A_{q'}(\omega_{0},\mu) \widetilde{\rho}(t) A_{q}(-\omega_{0},-\mu)\nonumber\\
- A_{q}(-\omega_{0},-\mu) A_{q'}(\omega_{0},\mu) 
\widetilde{\rho}(t)) + \emph{h.c.}
\label{360}
\end{eqnarray}
Then, using the matrix elements
\begin{eqnarray}
(A_q(\omega_{0},\mu ))_{nn'}=(A_q)_{nn'}\delta_{M'-M,\mu}
\delta(\epsilon_{0n'}-\epsilon_{0n},\hbar \omega_{0})~, 
\label{370}
\end{eqnarray}
one obtains the generalized Pauli equations as before
\begin{eqnarray}
\dot{\rho}_{nn} = \sum_{n'} W_{n'n}(t) \rho_{n'n'}-\sum_{n'} W_{nn'}(t) \rho_{nn}~,
\label{380} 
\end{eqnarray}   
where $W_{nn'}$ are given by
\begin{eqnarray}
W_{nn'}= \sum_{qq'} \gamma_{qq'}(\omega_{nn'}(t))(A_q)_{nn'}(A_{q'})_{n'n}~,
\label{390}
\end{eqnarray}
with $\gamma_{qq'}(\omega)$ as in Eq. (\ref{220}).
According to the above analysis, the dynamics of the reduced density matrix has the same major 
features as in the case of Section II, but now with the appropriate and reasonable modification 
for the excitation frequencies.

\begin{figure}[!t]
\centering
\includegraphics[width=3.4in, height=3in]{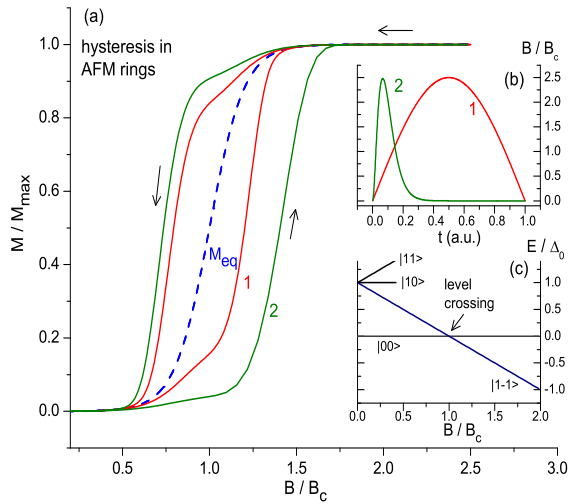}
\caption{(Color online) Hysteretic behavior 
of AFM ring systems at low $T$ and for two commonly used sweep forms $B(t)$ shown in Fig. 1(b). 
The loops are obtained by a numerical solution of 
Eq. (\ref{270}), with $M_{eq}$ given in the text and $1/\tau_{s}$ taken as the sum of the 
one-phonon term plus a small residual contribution.
All the parameters used are arbritrary. The equilibrium magnetization is denoted by the dashed line. 
The low-energy diagram and the true level crossing between the two relevant states $|0,0>$ and $|1,-1>$ 
is shown in Fig. 1(c).} 
\end{figure}
As discussed in the Introduction, a class of magnetic molecules where the above analysis 
can be easily applied is that of AFM rings at low $T$ and for fields in the vicinity of 
a given level crossing value. Specifically, we will assume $T<<\Delta_0/k_B$, where $\Delta_0$ denotes 
the first, zero-field excitation energy and consider fields in the vicinity of 
the first level-crossing field ($B_c=\Delta_{0}/\hbar\gamma$) only, where the singlet $|0,0>$ 
intersects the $M=-1$ (or $|1,-1>$) level of the $S=1$ triplet state (see Fig. 1(c));
a similar analysis can be employed for fields in the vicinity of higher level crossings. 
For these temperatures and fields, the AFM rings behave as a two-level system. 
Hence, we are dealing with a situation that is very similar to the spin 1/2 case,  
discussed in Sec. II. In fact, one can arrive at the same generalized Bloch equation 
for $M(t)$ (Eq. (\ref{270})), where now $M_{eq}=\hbar\gamma ~\textrm{sech}(\beta\Delta(t)/2)$, and 
$\Delta(t)\equiv \Delta_{0}-\hbar\gamma B(t)$. Similarly, $1/\tau_s$ will depend 
explicitly on $\Delta(t)$. For istance, the contribution $1/\tau_s^*$ of the one-phonon processes 
to the relaxation rate is given by Eq. (\ref{300}) with $\Omega(t)$ replaced by $\Delta(t)/\hbar$.

Similarly with the approach followed in Ref. \onlinecite{Rousochatzakis} for \{V$_6$\},
finding the physically appropriate functional form for $1/\tau_s$ 
can be facilitated by comparing the theoretical predictions with experimental data 
for a variety of field sweeps. 
We illustrate this idea in Fig. 1(a) by showing typical hysteresis loops 
obtained by numerically solving Eq. (\ref{270}) using
two different, commonly used sweeps shown in Fig. 1(b). The relaxation is assumed
to be driven by one-phonon processes plus a small residual term.
Since we have assumed an isotropic Hamiltonian and 
therefore a true level-crossing at $B\approx B_c$, nothing exciting happens in that regime. 
Instead, our choice for $1/\tau_s$ gives rise to a plateau for $B\approx B_c$
since in that regime the one-phonon term vanishes and the relaxation is driven only by the 
small residual term.
This is qualitatively different from observations in AFM ring systems such as    
the dimer \{Fe$_2$\},\cite{Shapira} and dodecanuclear $\{\textrm{Fe}_{12}\}$,\cite{Inagaki}  
which show a step-like behavior of $M(t)$ and consequently peaks in $dM(t)/dB$ at the level 
crossing fields. In addition, in some cases, the major peaks are accompanied with small satellite peaks;  
a small plateau in $M(t)$ is formed after each step (Foehn effect).
As we show in the following Section, both of these features, namely the magnetization steps and plateaus can 
be explained by extending our theory to include small off-diagonal terms in the spin Hamiltonian.  

\section{Dissipative LZS model in the adiabatic regime}
We now include small off-diagonal terms in the spin Hamiltonian 
in order to take into account the combined effect of LZS and thermal transitions.
We will consider the adiabatic regime: As mentioned in the Introduction, 
this is indeed the relevant regime for a large class of molecules
with low spin ground state, where the energy splittings are typically of order
$\delta/k_B \sim 0.1$ K. 
In particular, the magnetic molecules \{V$_6$\}\cite{Rousochatzakis} and  \{ V$_{15}$\}\cite{Chiorescu} 
as well as the AFM rings fall in this category. 
Since the effect of the small non-diagonal terms becomes manifest only in the immediate vicinity 
of the intersection of two\footnote{For \{V$_6$\}, as explained in Ref. \onlinecite{Rousochatzakis}, 
the number of intersecting energy levels at $B\approx 0$ is four.} energy levels, we are able to 
construct the following quite adequate theory. 

\subsection{General theory}
We consider two energy eigenstates of the isotropic Heisenberg Hamiltonian 
(denoted as $|m_1>$ and $|m_2>$) with total spins $S_1$ and $S_2$ 
and magnetic quantum numbers $m_1$ and $m_2$, respectively, 
which are coupled by a small off-diagonal (anisotropic) term. This term gives rise to 
an avoided level crossing between the two energy levels with a small energy gap denoted by $\delta$.
For fields in the immediate vicinity of this level anti-crossing, one can 
write the Hamiltonian in the basis of $|m_1>$ and $|m_2>$, i.e.,  
\be
H_s(t)=\left( \begin{array}{cc} 
E_0+m_1\hbar f(t) & \delta/2\\
\delta/2 & m_2\hbar f(t)
\end{array}
\right)~,
\label{420}
\ee
where $E_0$ denotes the zero-field energy difference between the two states for $\delta=0$.  
Without loss of generality, we assume that $m_2 > m_1$, with $E_0,\delta$ being
real and positive. In the absence of the off-diagonal term (i.e., $\delta=0$) the two levels cross at 
the moment when $f=f_c\equiv E_0/\hbar(m_2-m_1)$. 
We denote the adiabatic energy levels of $H_s(t)$ by $\epsilon_{\pm}(t)$ and the corresponding
eigenstates by $|\epsilon_{\pm};t>$, i.e.,
$H_s(t)|\epsilon_{\pm};t>=\epsilon_{\pm}(t)|\epsilon_{\pm};t>$. 
In the basis of $|m_1>$ and $|m_2>$, they can be expressed in the convenient parametric form 
\begin{eqnarray}
|\epsilon_{+};t>=\left(\begin{array}{c}
\textrm{cos}~\theta/2\\
\textrm{sin}~\theta/2
\end{array} \right)~, \nonumber\\ 
|\epsilon_{-};t>=\left(\begin{array}{c}
-\textrm{sin}~\theta/2\\
\textrm{cos}~\theta/2
\end{array}\right)~,
\end{eqnarray}
and
$\epsilon_{\pm}(t)=[E_0+\hbar f(m_1+m_2) \pm \hbar\Omega]/2$, where 
\be
\Omega\equiv[(\delta/\hbar)^2+(m_1-m_2)^2(f-f_c)^2]^{1/2}. 
\ee
The time-dependent parameter $\theta$ is given by 
\be
\theta=\textrm{tan}^{-1} \frac{\delta/\hbar}{(m_1-m_2)(f-f_c)}~, 
\ee
and extends from $0$ to $\pi$ as $f$ goes fron $-\infty$ to $+\infty$. 
In particular, $\theta \approx \pi/2$ when $f$ is in the immediate vicinity of $f_c$.

As in Secs. II and III, we switch to the interaction picture
\be
\widetilde{\rho}\equiv U_s^{\dagger}\rho U_s ~,
\label{450}
\ee
where $U_s(t)$ is the evolution operator
for the spin Hamiltonian $H_s(t)$ alone, which obeys $i\hbar\dot{U}_s = H_s(t)U_s$.  
Contrary to the previous isotropic cases, it is clear that 
due to the presence of two non-commuting terms in $H_s(t)$ the form of $U_s(t)$ cannot
be written in a closed analytical form.\footnote{Compactly written, $U_s(t)=T e^{-i\int_{-\infty}^{t}dt' H_s(t')}$, 
where $T$ denotes the chronological operator.} 
Thus, an analysis parallel to that of the previous sections cannot be 
readily employed. Nevertheless, it is possible to circumvent this difficulty by exploiting the fact that we 
are in the adiabatic regime. According to the adiabatic theorem one has   
\be
U_s(t)|\alpha;-\infty>=e^{-i\phi_{\alpha}(t)}|\alpha;t>,
\label{460}
\ee
where the phases $\phi_\alpha(t)$ are given by
$\phi_{\alpha}(t)=\int_{-\infty}^{t}\epsilon_{\alpha}(t')dt'/\hbar$. 
As we mentioned before, adiabaticity holds even inside the LZS regime.

We now express the spin operators $A_q$ appearing in the spin-phonon interaction term $V$ (Eq. (\ref{20})), 
in the adiabatic basis as
\be
A_q=\sum_{\alpha,\beta} A_q^{\alpha\beta}(t) |\alpha;t><\beta;t|,
\label{470}
\ee
where $A_q^{\alpha\beta}(t) \equiv <\alpha;t|A_q|\beta;t>$. 
In the interaction picture these take the form
\be
\widetilde{A}_q(t)=\sum_{\alpha,\beta}A_q^{\alpha \beta}(t) e^{i(\phi_{\alpha}(t)-\phi_{\beta}(t))}
|\alpha;-\infty> <\beta;-\infty| ~.
\label{480}
\ee
Following the previous steps, it is straightforward
to derive an expression similar to Eq. (\ref{120}), where one encounters  
typical matrix elements such as $A_q^{\alpha\beta}(t-u)$ 
and phase factors of the form $e^{i\phi_{\alpha}(t-u)}$. 
The next step is to approximate $A_q^{\alpha\beta}(t-u)\approx A_q^{\alpha\beta}(t)$, and 
also to make a Taylor expansion of the phases in first order in $u$, as before, i.e.,  
\be
\phi_{\alpha}(t-u)\approx \phi_{\alpha}(t)-u ~\epsilon_{\alpha}(t)/\hbar.
\label{490}
\ee 
It is then possible to obtain the following set of Pauli master equations by introducing the representation  
$\widetilde{\rho}_{\alpha\beta}(t) \equiv <\alpha;-\infty|\widetilde{\rho}(t)|\beta;-\infty>$, 
and then performing the RWA,  
\be
\dot{\widetilde{\rho}}_{\alpha\alpha}=W_{\beta\alpha}(t)\widetilde{\rho}_{\beta\beta}-
W_{\alpha\beta}(t)\widetilde{\rho}_{\alpha\alpha},
\label{510}
\ee
where the transition rates $W_{\alpha\beta}$ are now given by
\be
W_{\alpha\beta}(t)=\sum_{qq'} A_q^{\alpha\beta}(t) A_{q'}^{\beta\alpha}(t)\gamma_{qq'}(\omega_{\alpha\beta}(t)),
\label{520}
\ee
and $\gamma_{qq'}(\omega)$ as in Eq. (\ref{220}).
We now compare this expression for the transition rates with the corresponding 
ones found previously. The new ingredient here is the extra time dependence of the transition rates 
carried by the matrix elements $A_q^{\alpha\beta}(t)$. 
Clearly this results from the explicit time dependence of the adiabatic energy states. 
Physically this means, for example,
that longitudinal fluctuating fields (contained in $V$) can become effective in inducing transitions 
between the two levels inside the LZS regime
because of the admixing of the two states. 
This introduces an additional complication when one attempts to quantitatively account for 
magnetization data inside the LZS regime. 
\begin{figure}[!t]
\centering
\includegraphics[width=3.4in, height=3in]{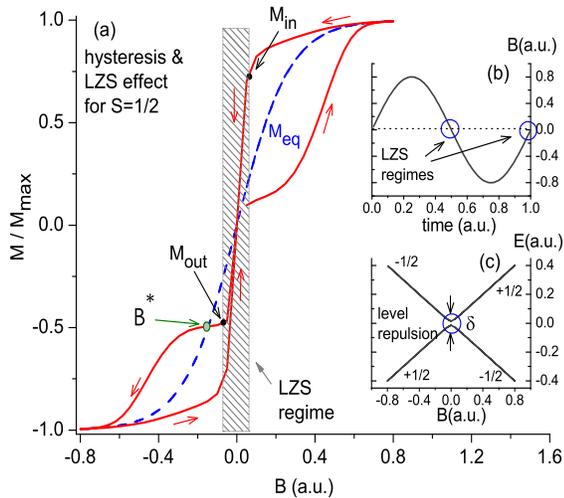}
\caption{(Color online) Hysteresis and LZS effect at $B\approx 0$ 
for the case of an S=1/2 magnetic molecule, as obtained by solving numerically our equations (see text) 
for the full-cycle sweep shown in Fig. 2(b). 
The shaded area indicates the LZS regime.
The equilibrium magnetization is denoted by the dashed line.
Note the deviation from the exact magnetization reversal at $B\approx 0$ as described in the text, 
and the formation of the small plateau after the step. 
$B^*$ denotes the field at which $M$ crosses $M_{eq}$, as discussed in the text.
The low-energy diagram and the level anti-crossing between the two levels $|-1/2>$ and 
$|+1/2>$ is shown in Fig. 2(c). The small circles in Figs. 2(b) and 2(c) 
indicate the LZS regimes, where the magnetization steps take place.} 
\end{figure} 

It turns out that one can obtain a simplified 
Bloch type of equation for the quantity $n(t)\equiv\widetilde{\rho}_{--}-\widetilde{\rho}_{++}$, 
in terms of which the magnetization can be simply expressed.
Using the above master equations and the normalization condition 
$\widetilde{\rho}_{--}+\widetilde{\rho}_{++}=1$, one arrives at the following equation for $n(t)$
\be
\dot{n}(t)=\frac{1}{\tau_s(t)}[n_{eq}(t)-n(t)] ~,
\label{530}
\ee
where $n_{eq}=\textrm{tanh}(\beta\hbar\Omega/2)$, 
$1/\tau_s=W_{-+}\left(1+e^{\beta\hbar\Omega}\right)$. 
In order to express $<S_z>=\textrm{Tr}\{\rho S_z\}$ in terms of $n(t)$ one notes that
\begin{eqnarray}
<S_z>=\sum_{\alpha,\beta}<\alpha;t|\rho|\beta;t> S_z^{\beta\alpha}(t) \nonumber\\
=\sum_{\alpha,\beta}e^{i(\phi_{\beta}(t)-\phi_\alpha(t))}\widetilde{\rho}_{\alpha\beta}
S_z^{\beta\alpha}(t) ~.
\label{540}
\end{eqnarray}
Now, for the longitudinal magnetization one can safely repeat the RWA by neglecting the terms 
with $\alpha\ne\beta$. Then $<S_z>\approx \widetilde{\rho}_{--} S_z^{--}+
\widetilde{\rho}_{++} S_z^{++}$, or for the magnetization $M$,
\be 
M\approx\frac{-\hbar\gamma}{2}[(m_1+m_2)-(m_2-m_1)^2 n(t)\frac{f-f_c}{\Omega}]~.
\label{550}
\ee
Equations (\ref{530}) and (\ref{550}) are of central importance 
for describing the combined effects of LZS and thermal transitions.\footnote{It is noted that 
the analysis of Ref.\onlinecite{Saito}, which is based on a quantum master equation previously 
developed for static fields, provides an alternative set of phenomenological equations, 
to be contrasted to Eqs. (\ref{530}) and (\ref{550}) that we derived from first-principles.}  
According to Eq. (\ref{550}), the magnetization is given in terms of two distinct quantities, 
namely the ratio $(f-f_c)/\Omega$ and $n(t)$. 
Interestingly, the factor $(f-f_c)/\Omega$, carries the physics of the purely quantum LZS
effect, since it changes sign inside the LZS regime as expected in the adiabatic regime of
the LZS transitions. On the other hand, the quantity $n(t)$ contains
all information about thermal transitions and dissipation since its dynamics is determined by the
spin-bath coupling according to Eq. (\ref{530}). 

\subsection{Qualitative analysis}
We will now give a qualitative analysis of the predictions of this theory 
by examining the general structure of Eqs. (\ref{530}) and (\ref{550}). We will also 
demonstrate the main ideas by providing numerical solutions for typical examples, namely 
the $S=1/2$ case and that of AFM rings at low $T$.
These are shown in Figs. 2, 4 and in Fig. 3, respectively. 
For these solutions, we have assumed that $1/\tau_s$ is the sum of the one-phonon process 
contribution plus a small residual term to account for the relaxation inside the LZS regime. 
The parameters chosen are somewhat arbritrary;
the exact shape of the $M$ vs $B$ curves is determined by the actual parameters of a
given system (magnitude of $\delta$, spin-phonon coupling terms, etc). 
As explained below, Eqs. (\ref{530}) and (\ref{550}) account nicely for 
all the dynamical effects shown in pulsed field measurements, namely hysteresis loops (already discussed in 
Secs. II and III) outside the LZS regime, 
the thermal deviation of the LZS steps from the pure quantum-mechanical prediction 
(see in particular Figs. 2 and 3), as well as the formation of magnetization plateaus (Foehn effect) 
immediately after exiting the LZS regime (see Figs. 2 and 3, but mostly Fig. 4). Furthermore, we will 
discuss how the PB effect, which takes place at very low $T$, can give rise to an enhancement 
of the Foehn effect.    

\subsubsection{Hysteresis loops}
We begin by noting that for either $\delta=0$ or for fields outside the LZS regime 
one has $(f-f_c)/\Omega\propto \textrm{sgn}(f-f_c)$ and therefore the only time dependence of $M$ 
stems from the quantity $n(t)$. In addition, all matrix elements $A_q^{\alpha\beta}$ appearing in Eq. (\ref{520}) 
become time-independent. Then, by taking the time derivative of Eq. (\ref{550}),
one recovers the results for the isotropic case, and in particular the 
generalized Bloch equation, Eq. (\ref{270}), derived before for the case of $S=1/2$ and that of the AFM 
rings at low $T$. Thus, as expected, one can neglect the LZS effect 
for fields outside the immediate vicinity of level crossings. This justifies 
the use of Eq. (\ref{270}) in Ref. \onlinecite{Rousochatzakis}, for fields away from $B\approx 0$. 
Typical hysteresis loops for fields outside the LZS regime 
are shown in Figs. 2(a) and 3(a) for the $S=1/2$ case and that of AFM rings at low $T$, 
respectively. One should note that
although the magnetization obeys the same generalized Bloch equation (Eq. (\ref{270})) 
for either $\delta=0$ or for $\delta \ne 0$ but for fields outside the LZS regime, the solution is drastically 
different for these two cases (compare for example Figs. 1 and 3, for the case of AFM rings). 
This is because, when $\delta \ne 0$, 
the occurence of an LZS step introduces a different (as compared to the $\delta=0$ case) 
initial condition immediately after exiting the LZS regime. A direct consequence
of this is the Foehn effect discussed below.  

\subsubsection{Thermal corections to the LZS step}
A deviation from the exact quantum-mechanical prediction regarding the magnetization step
at $f\approx f_c$, is expected to arise from thermal transitions inside the LZS regime.
This can be seen as follows. Assuming that one crosses $f_c$ from below, and
denoting by $M_{in}$ and $M_{out}$ the magnetization when entering 
and when exiting the LZS regime, 
respectively, (similarly for $n_{in}$ and $n_{out}$), we can obtain from Eq. (\ref{550}) 
\be
M_{in}+M_{out}=-\hbar\gamma [(m_1+m_2)-(m_2-m_1)^2\delta n_{LZS}/2]~,
\label{560}
\ee
where $\delta n_{LZS}\equiv n_{out}-n_{in}$, denotes the overall change of $n(t)$ inside the LZS 
regime, as obtained using Eq. (\ref{530}). This quantity is negative, i.e., $n_{out} < n_{in}$ 
(this can be easily seen by plotting $n_{eq}$ vs $f$ 
and solving Eq. (\ref{530}) graphically, i.e., without specifying the form of $1/\tau_s$). 
The first term of Eq. (\ref{560}) gives the quantum-mechanical prediction in the adiabatic regime, 
since in the absence of thermal transitions inside the LZS regime (i.e., $\dot{n}(t)=0$) the second term vanishes.
To be more specific, for the spin $S=1/2$ case ($m_2=-m_1=1/2$, and $E_0=0$), Eq. (\ref{560}) gives 
\be
M_{out}=-M_{in}-\hbar\gamma |\delta n_{LZS}|/2~.
\label{565}
\ee
One then obtains the expected magnetization reversal ($M_{out}=-M_{in}$) in the absence of
thermal effects ($\delta n_{LZS}=0$).  
Thus, the second term of Eq. (\ref{560}) (or that of Eq. (\ref{565})) 
gives the thermal correction; its magnitude clearly depends on the competition between two 
time scales, namely $\tau_s$ and the time $\delta t_{LZS}$ spent inside the LZS regime, given by 
$\delta t_{LZS}=2\delta/(\hbar\gamma r)$, which is controlled by the sweep rate $r$. 
Hence, for a given $T$, the thermal correction becomes larger with 
decreasing sweep rates.\footnote{This behavior should be contrasted with 
the sweep rate dependence of the magnetization steps
for the pure quantum-mechanical LZS model but in the non-adiabatic regime, which is the relevant case for SMM's. 
Acording to the well known formula $P_{LZS}=1-\textrm{exp}[-\pi \delta^2/(2 \hbar^2 \gamma r)]$ for 
the transition probability $P_{LZS}$, when lowering $r$ one increases $P_{LZS}$ i.e., reduces 
the effect of non-adiabatic transitions, thus giving rise to a larger height of the magnetization step.}
Furthermore, since $1/\tau_s$ is expected to increase with increasing $T$ 
the thermal effects are more pronounced at higher $T$, as 
indeed observed for \{V$_6$\}.\cite{Rousochatzakis}
Of course, at high enough $T$, the LZS effect is completely masked by the thermal transitions 
and the step disappears. 
More generally, it should be noted that it is $M_{in}+M_{out}$, rather than
$M_{out}-M_{in}$ (height of the step), that is of more direct
relevance in experimentally determining the extent of the thermal effects.   
The numerical solution for the $S=1/2$ case
shown in Fig. 2(a) demonstrates the deviation from exact magnetization reversal at 
$B\approx 0$, as a result of thermal transitions inside the LZS regime. 
This is consistent with the experimental
data for \{V$_6$\}. 
\begin{figure}[!t]
\centering
\includegraphics[width=3.4in, height=3in]{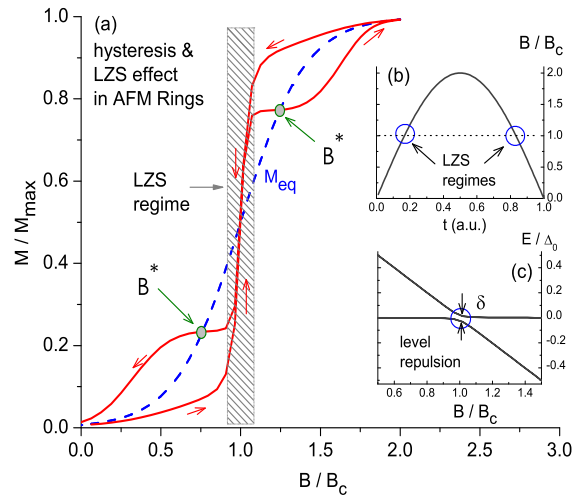}
\caption{(Color online)  Hysteresis and LZS effect at $B\approx B_c$ 
for AFM ring systems at low $T$ and for fields around the first level-crossing value $B_c$, 
as obtained by solving numerically our equations (see text) for the half-cycle sweep shown in Fig. 3(b). 
The shaded area indicates the LZS regime.
The equilibrium magnetization is denoted by the dotted line.
Note the deviation from the pure quantum mechanical prediction regarding the step  
(see text) and the formation of small plateaus 
after each step. $B^*$ denotes the fields at which $M$ crosses $M_{eq}$, as discussed in the text. 
The low-energy diagram and the level anti-crossing between the levels $|0,0>$ and 
$|1,-1>$ is shown in Fig. 3(c). The small circles in Figs. 3(b) and 3(c) indicate the LZS regimes, where
the magnetization steps take place. This typical behavior of $M(t)$ in the case of $\delta \ne 0$
should be contrasted with that of $\delta=0$ shown in Fig. 1.} 
\end{figure} 
\subsubsection{The magnetic Foehn effect}
Another effect which is also of particular interest is the ``magnetic Foehn effect''\cite{Saito} 
mentioned in the Introduction, which concerns the formation of plateaus shortly after 
exiting the LZS regime. Although these plateaus occur outside the LZS regime, 
they are a direct concequence of the LZS effect, as can easily be explained by considering the
particular example shown in Fig. 4 (the analysis for the other cases shown in Figs. 2(a) and 3(a) 
is analogous). Immediately after exiting the LZS regime, $M_{out} > M_{eq}$ in contrast to the  
typical hysteretic behavior where $M$ lags behind $M_{eq}$. 
Then, outside the LZS regime where, as explained above, $M$ tends towards $M_{eq}$, 
$M$ will decrease even though the field is swept towards larger values.
This drop of $M$ eventually stops at a field $B^*$, at the moment when $M$ equals $M_{eq}$. 
Thus, $B^*$ corresponds to a local minimum in $M$(since then $\dot{M}=0$). 
Generally, this minimum will be observable if $\tau_s$, in 
that regime, is much shorter than the sweep time $\tau_e$. 
If this is not the case, a broad plateau around $B^*$ is formed instead. In fact, this seems (see below) to be
the typical behavior in all experimental data reported so far.  
This behavior is shown in Fig. 4(a), where for a fixed $T$ and with increasing sweep rate $r$, the minimum broadens
with $B^*$ departing from $B_c=0$. One can exploit this feature experimentally, 
in order to obtain an estimate for $\delta$, since all $B^*$ corresponding to different sweep rates must lie 
on the equilibrium curve $M_{eq}(B(t))$.  

The curves shown in Fig. 4 also suggest why  a very steep minimum at $B^*$ (pronounced Foehn effect) 
has not been reported so far in realistic situations. In principle, as mentioned above, the minimum at $B^*$ should 
become more steep at lower sweep rates. On the other hand, on decreasing the sweep rate, one indirectly 
reduces the value of $M_{out}$,\footnote{The reduction 
of $M_{out}$ is due to thermal dissipation before 
entering the LZS regime (which determines the value of $|M_{in}|$ and consequently 
$M_{out}$ through Eq. (560)) as well as inside the LZS regime (shaded area in Fig. 4).}  
thus being closer to $M_{eq}$ when exiting the LZS regime, 
and therefore compensating the effect of slowing down the sweep rate.  

Ideally, for the realization of a very steep minimum at $B^*$ one needs 
(for a fixed sweep rate), a relaxation process that is slow enough at negative fields 
(i.e., large $|M_{in}|$ and concequently 
large $M_{out}$) but becomes faster at positive fields, after exiting the LZS regime. 
Such an ``asymmetry'' in the field dependence of $1/\tau_s$ cannot exist 
unless we abandon our main assumption that the phonons are in equilibrium at all experimental times.
This is because, according to our analysis, with the phonons being in equilibrium, 
$1/\tau_s$ depends on $|B(t)|$, and thus must be symmetric around $B_c=0$.  
This raises the question whether the PB effect, which takes place at very low $T$, could give rise to 
a pronounced Foehn effect.  
This is indeed supported by the analysis of the PB effect described 
in Ref. \onlinecite{Chiorescu}, for the magnetic molecule \{ V$_{15}$\}, which shows an asymmetry in the 
number of resonant phonons before entering and after exiting the LZS regime 
(see, in particular, the inset of Fig. 2(b) of Ref. \onlinecite{Chiorescu}), as a result of the strong coupling 
of the phonons to the spin degrees of freedom.
In short, the physical origin of this asymmetry is that $B_c=0$ is a level-crossing field for 
the energy spectrum of the spin system, whereas it corresponds to a ``reflection'' point 
for that of the resonant phonons: Although, in general, when sweeping an external field one progressively 
brings the spin system into resonance with different phonon modes (instantaneous resonance condition) 
this is not true in the vicinity of a level anti-crossing, since when exiting the LZS regime 
the relaxation is driven by the same phonons that were in thermal contact with the spins 
while entering the LZS regime (for more details, see Refs. \onlinecite{Chiorescu,Waldmann}). 
The above enhancement of the Foehn effect due to the PB effect has been in fact recently observed 
experimentally and will be reported elsewhere.\cite{Wernsdorfer2}
\begin{figure}[!t]
\centering
\includegraphics[width=3.4in, height=3in]{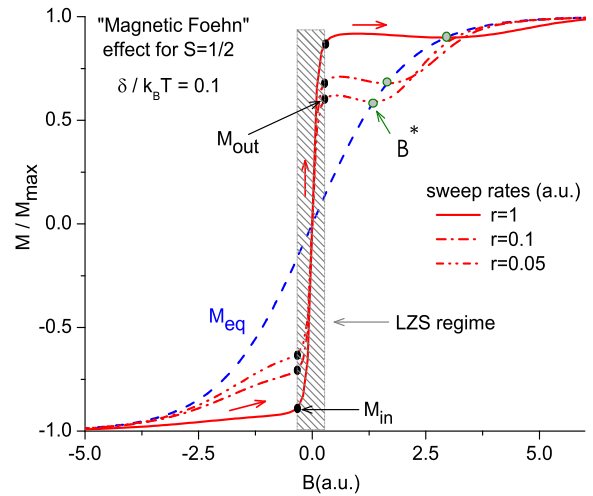}
\caption{(Color online) Magnetic Foehn effect  
for a magnetic molecule with a spin $S=1/2$ ground state, for fixed $T$ and for three different fields sweeps 
with constant sweep rates $r$ (in arbritrary units). The shaded area denotes the LZS regime.
Because of the LZS effect, immediately after exiting the LZS regime (shaded area)
$M_{out} > M_{eq}$, as opposed to a usual hysteresis situation. Then, 
outside the LZS regime, where $M$ tends towards $M_{eq}$ (see text),
 $M$ drops even though the field increases; a local minimum of $M$ is formed at a field $B^*$ 
at which $M=M_{eq}$. Note that this minimum is departing from the LZS regime of the magnetization step 
and broadens (eventually giving rise to a plateau) with increasing sweep rates.} 
\end{figure} 

\section{Summary}
We have extended the standard spin-lattice relaxation theory, in the context
of pulsed field studies of magnetic molecules. 
Being easily applied to simple systems, this generalized theory can give 
important information on the underlying relaxation mechanisms and the microscopic interactions present 
in magnetic molecules in general. 
All the dynamical magnetization effects, including hysteresis loops, 
LZS transitions with or without 
dissipation and 
magnetization plateaus (Foehn effect), which are manifested in pulsed fields measurements, 
are accounted for by the comprehensive theory presented above.

We first developed the theory for the isotropic case of molecules with a spin $S$ ground state
well separated from the excited levels but also the general Heisenberg model where all energy levels 
are relevant. 
We have shown, for two such simple cases, namely the spin $S=1/2$ case and that of 
AFM ring systems at low $T$, that the dynamical magnetization obeys a 
generalization of the standard Bloch equation.
In particular, this equation has been recently used\cite{Rousochatzakis}
for the magnetic molecule \{V$_6$\}, and was found to provide results in excellent 
agreement with experimental data at $T>1$ K, confirming that the dominant mechanism driving the relaxation is
the one-phonon processes, and in addition providing an estimate of the spin-phonon coupling energy. 
Obtaining this information was greatly facilitated by using a variety of field sweep forms. 

We also extended the theory to include small off-diagonal terms in the spin 
Hamiltonian and thus take into account the combined effects of LZS and thermal transitions. 
This was done here for the large class of magnetic molecules with a low spin ground state.
For these molecules, and for the currently available sweep rates, one is in the extreme 
adiabatic regime.   
Our main interest in these systems, has been the role of dissipation on the LZS steps, 
as well as the formation of small plateaus (Foehn effect) formed after each step,
observed in several magnetic molecules. 
Interestingly enough, we arrived at a convenient set of equations where the effects of dissipation
and LZS transitions can be treated separately. 
These equations account nicely for the description of both the magnetization steps and the plateaus. 
Moreover, the role of temperature and the field sweep rate on these effects becomes transparent.
Finally, although our analysis is limited to high enough temperatures ($T\gtrsim 1$ K) so that 
one can assume that the phonons are in equilibrium at all experimental times,
it nevertheless indicates how an enhanced Foehn effect could arise at lower $T$, where 
the PB effect takes place. In such a case, steep extrema in the $M$ vs $B$ curves 
could be observed. 

The present theoretical work provides a first step towards exploiting the possibilities that are
offered by probing magnetic molecules using external magnetic fields with high sweep rates. These probes,
apart for providing information specific to magnetic molecules, offer the possibility of conducting a detailed 
study of the relaxational behavior of interacting spin systems as a result of their coupling with a ``heat bath''
and in particular the excitations of the host lattice. Development of a broad theoretical framework for dealing 
with relaxational phenomena induced by dynamical magnetic fields is indeed a worthy goal.   

\section{Aknowledgments}
We thank Prof. Yoshitami Ajiro (Kyoto University) for very helpful comments that
have greatly improved our manuscript.
Ames Laboratory is operated for the U.S. Department of Energy by Iowa State University under contract 
No. W-7405-Eng-82.

\end{document}